# Stacking an autoencoder for feature selection of zero-day threats

## Mahmut TOKMAK*[1], Mike NKONGOLO[2]

**Abstract**: Zero-day attacks exploit previously unknown vulnerabilities in software, hardware, or networks. Since these vulnerabilities are not yet patched or protected against, they present a significant risk. Detecting zero-day attacks is crucial due to their exploitation of unknown vulnerabilities, posing significant risks to cybersecurity. Timely detection allows for swift response and deployment of countermeasures, minimizing the window of opportunity for attackers. It protects sensitive data, mitigates financial losses, safeguards reputation, and strengthens cybersecurity practices. Detecting zero-day attacks provides insights into attack vectors, improves security measures, and enhances incident response capabilities. It helps prevent future attacks by understanding adversary techniques and updating defense mechanisms. Overall, zero-day attack detection plays a critical role in mitigating risks, protecting assets, and staying ahead in the evolving threat landscape. This study explores the application of stacked autoencoder (SAE), a type of artificial neural network, for feature selection and zero-day threat classification using a Long Short-Term Memory (LSTM) scheme. The process involves preprocessing the UGRansome dataset and training an unsupervised SAE for feature extraction. Fine-tuning with supervised learning is then performed to enhance the discriminative capabilities of this model. The learned weights and activations of the autoencoder are analyzed to identify the most important features for discriminating between zero-day threats and normal system behavior. These selected features form a reduced feature set that enables accurate classification. The results indicate that the SAE-LSTM performs well across all three attack categories by showcasing high precision, recall, and F1 score values, emphasizing the model's strong predictive capabilities in identifying various types of zero-day attacks. Additionally, the balanced average scores of the SAE-LSTM suggest that the model generalizes effectively and consistently across different attack categories. The SAE-LSTM model excels in detecting signature attacks, while synthetic signature and anomaly attacks pose challenges due to abnormality or absence of patterns. The methodology we put forth utilizes the SAE-LSTM technique, resulting in a remarkable accuracy of 98%, outperforming prior intrusion detection studies. Hence, this research aims to contribute to advanced cyberintelligence to proactively mitigating zero-day threats. Future cyberintelligence can progress by refining the proposed feature selection model, addressing computational efficiency of the UGRansome dataset, and integrating hybrid methodologies for improved intrusion detection capabilities.

**Keywords**: stacked autoencoder, feature selection, zero-day threats, machine learning, deep learning, UGRansome, cyberintelligence

[1]**Address:** Burdur Mehmet Akif Ersoy University, Bucak Zeliha Tolunay School of Applied Technology and Management, Burdur/Turkiye
[2]**Address:** University of Pretoria, Faculty of Engineering, Built Environment and Information Technology, Department of Informatics, Pretoria/South-Africa

*****Corresponding author**: mahmuttokmak@mehmetakif.edu.tr

## 1. INTRODUCTION

In the ever-evolving landscape of cybersecurity, the emergence of advanced and elusive threats poses unprecedented challenges to organizations, governments, and individuals alike. Among these threats, zero-day attacks have garnered substantial attention due to their potential to exploit undiscovered vulnerabilities and wreak havoc on digital infrastructure. A zero-day attack refers to a cyber assault that targets a previously unknown vulnerability in software, hardware, or network systems. These vulnerabilities, referred to as "zero-day vulnerabilities," are named as such because developers have "zero days" to address and patch them before malicious actors exploit them. Zero-day attacks are particularly effective because they can strike unexpectedly and catch defenders off guard. This surprise factor lets them evade regular security measures. Unlike known vulnerabilities, zero-day vulnerabilities have no documented fixes or protections, giving attackers a notable upper hand. Consequently, zero-day attacks present a daunting challenge for defenders, as they often evade signature-based detection systems and intrusion prevention tools

(Nkongolo & Tokmak, 2023; Thomas et al., 2021; Tokmak, 2022). In the era of big data, extracting meaningful and representative features from high-dimensional datasets has become a cornerstone of modern data analysis and machine learning (Nkongolo et al., 2022). Among the myriad of techniques, stacked autoencoders (SAEs) have emerged as a powerful tool for automated feature learning, enabling the discovery of intricate data structures and patterns. Rooted in the field of deep learning (DL), SAEs offer a compelling solution to the challenge of high-dimensional data representation, presenting a pathway towards enhanced predictive modeling, efficient dimensionality reduction, and insightful data interpretation (Boussaad & Boucetta, 2021; Kim et al., 2020). Feature selection with SAEs has been explored in several papers and various domains.

Wang et al. proposed the broad autoencoder features (BAF) which consist of four parallel interconnected SAEs with different activation functions (T. Wang et al., 2021). Kong et al. used a stacked autoencoder with L21-norm regularization for dimensionality reduction and feature extraction from electricity load data (Kong et al., 2020). Wang et al. introduced a stacked supervised auto-encoder (SSAE) to acquire fault-relevant attributes from raw input data and improve fault classification accuracy (Y. Wang et al., 2020). Chatterjee et al. integration of SAE characteristics with wavelet-based and morphological fractal texture attributes for the classification of skin disorders achieved high accuracy (Chatterjee et al., 2019). Kim et al. suggested an enhancement in tool condition diagnosis through the utilization of SAE-based CNC machine tool prognosis, incorporating feature extraction from discrete wavelet transform (Kim et al., 2020). Zero-day threat detection involves identifying and mitigating attacks that exploit unknown vulnerabilities, such as a cybercriminal targeting a newly discovered weakness in a popular software program before the software developer has a chance to release a patch. Nkongolo et al. introduced the UGRansome dataset, which facilitates the identification of anomalies and zero-day attacks that cannot be recognized by known threat signatures (Nkongolo et al., 2021). Kumar and Sinha proposed a resilient and smart cyber-attack detection model that uses the notion of prominent entities and network structure techniques to detect zero-day attacks (Kumar & Sinha, 2021). Sarhan et al. proposed a zero-shot learning approach for assessing the effectiveness of machine learning models in detecting zero-day attacks (Sarhan et al., 2023).

Millar et al. suggested a deep neural network for Android malware detection without prior knowledge of malicious characteristics, achieving high detection rates for zero-day scenarios (Millar et al., 2021). Blaise et al. proposed the utilization of the Split-and-Merge technique to promptly identify emerging botnets and recently exploited vulnerabilities, reducing false positives (Blaise et al., 2020). Long Short-Term Memory (LSTM) networks have displayed significant potential in unknown vulnerabilities and malware detection. Researchers have dedicated significant efforts to studying LSTM hyperparameters for the development of Intrusion Detection Systems (IDS). They have explored diverse LSTM setups and configurations. What they have discovered is that the significance of these hyperparameters in the context of IDS varies from their importance in language model applications. The interplay of hyperparameters significantly influences the assessment of their respective significance. Considering this interaction, the precise order of importance for LSTMs in IDSs includes batch size as the most critical, followed by dropout ratio and padding. Additionally, sensitivity-based LSTM models have been proposed for designing System-call Behavioral Language (SBL) for malware detection. These models achieve impressive Area Under the Curve (AUC) values and specificity on unknown attack datasets.

Another approach involves using LSTM with word embedding and attention mechanisms to effectively represent and classify malware files, achieving high accuracy and F1 scores (Sewak et al., 2021; Q. Xie et al., 2020; W. Xie et al., 2020; Zhang, 2020). A method for zero-day detection using LSTM is proposed in the paper by Fang et al. The model is designed to detect malicious JavaScript code injected into web pages. It extracts features from the semantic level of bytecode and optimizes the method of word vector. The LSTM-based detection model outperformed existing models based on Random Forest and Support Vector Machine (SVM) algorithms (Fang et al., 2018). Another paper by Roberts and Nair introduces a neural architecture for anomaly detection in discrete sequence datasets. Their model combines a modified LSTM autoencoder with an array of One-Class SVMs to find anomalies within sequences. The proposed method shows improved stability and outperforms standard LSTM and sliding window anomaly detection systems (Roberts & Nair, 2018).

This study aims to leverage the synergistic capabilities of SAE and LSTM networks to improve the identification and categorization of zero-day threats using the UGRansome dataset. The primary objective is to integrate feature selection techniques into the SAE architecture to streamline the extraction of pertinent and differentiating features from raw data. By carefully choosing the input data, the subsequent LSTM network can adeptly capture temporal relationships within the feature domain. Ultimately, this research strives to advance proactive and resilient cybersecurity strategies by introducing an innovative approach – a feature-selection-driven SAE-based LSTM model. The subsequent sections of this work delve into the methodology, experimental setup, results, and discussions, all of which culminate in a comprehensive analysis of the proposed SAE-based LSTM model for zero-day threats detection using the UGRansome dataset.

## 2. MATERIAL AND METHOD

### 2.1. Experimental Dataset

In 2021, Nkongolo et al. (Nkongolo et al., 2021) introduced the UGRansome dataset, a novel and comprehensive anomaly detection dataset designed to detect unknown network attacks, including zero-day threats (Nkongolo & Tokmak, 2023). Unlike existing datasets in the IDS field, UGRansome includes previously unexplored unknown and ransomware attacks (Nkongolo et al., 2022). The dataset comprises various attack categories, namely Signature (S), Anomaly (A), and Synthetic Signature (SS) (Figure 1), each with labeled instances of zero-day threats such as Locky, CryptoLocker, advanced persistent threats (APT), SamSam, WannaCry, Razy, JigSAW, Globe, TowerWeb, and more (Figure 1).

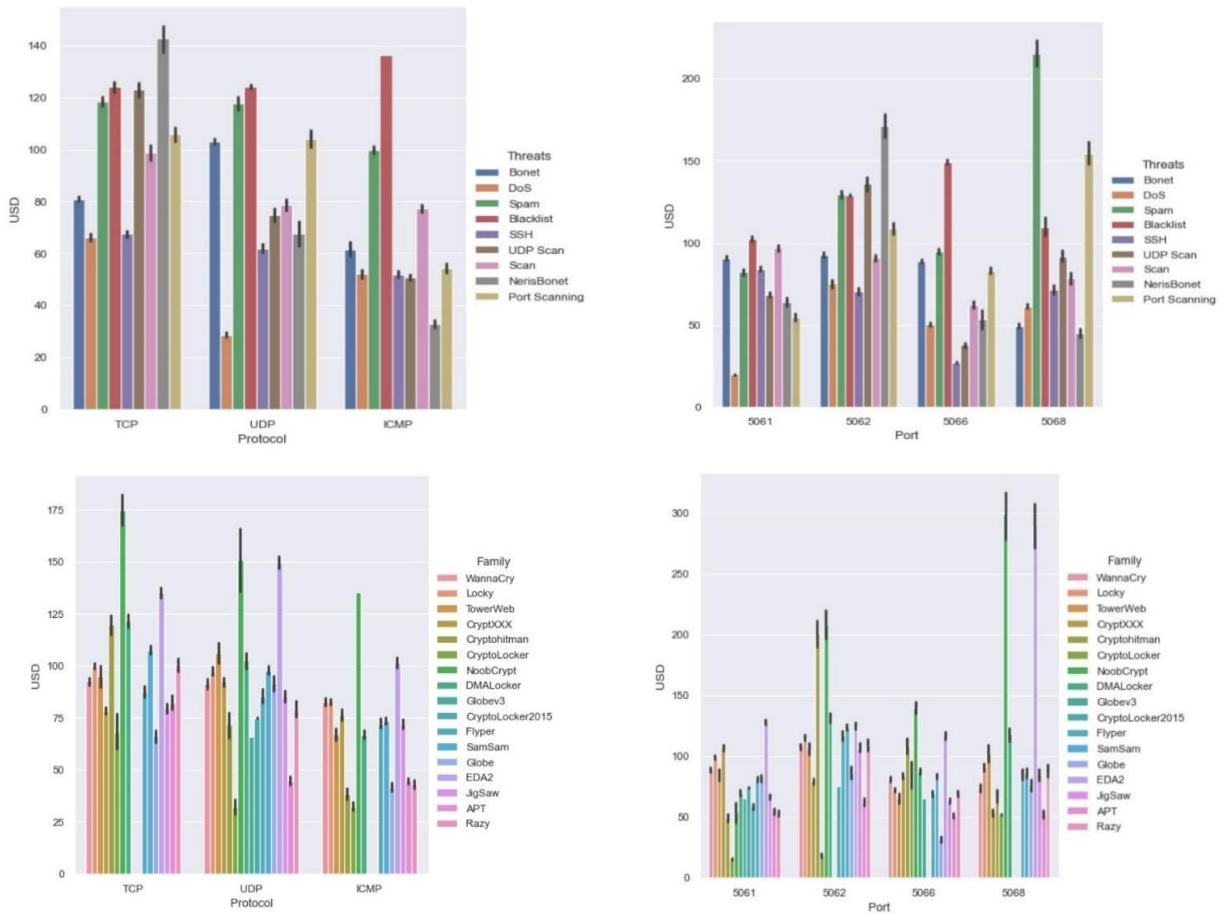

Figure 1. Distribution of zero-day threats categories of the UGRansome dataset

Table 1 provides an overview of the UGRansome features (f1-f18), while Table 2 displays the distribution of the UGRansome subsets. To gain a comprehensive understanding of the dataset statistics, we direct readers to Figure 2, which presents its key characteristics.

Table 1. The data structure of the UGRansome dataset

| No | Name | Type | No | Name | Type |
|----|------|------|----|------|------|
| f1 | SS | Categorical | f2 | Cluster | Numeric |

| f3 | S | Categorical | f4 | A | Categorical |
|---|---|---|---|---|---|
| f5 | Spam | Categorical | f6 | BTC | Numeric |
| f7 | Blacklist | Categorical | f8 | Bytes | Numeric |
| f9 | Nerisbotnet | Categorical | f10 | USD | Numeric |
| f11 | UDP scan | Categorical | f12 | JigSAW | Categorical |
| f13 | SSH | Categorical | f14 | Port | Numeric |
| f15 | DoS | Categorical | f16 | CryptoLocker | Categorical |
| f17 | Port scanning | Categorical | f18 | WannaCry | Categorical |

Table 2. The UGRansome subsets

| Dataset | A | S | SS |
|---|---|---|---|
| UGRansome19Train | 40.323 | 25.822 | 9.656 |
| UGRansomeVal | 11.869 | 9.439 | 1.736 |
| UGRansome18Test | 4.701 | 3.408 | 6.954 |
| Total | 56.893 | 38.669 | 18.346 |
| Average (avg) | 19.964 | 12.889 | 6.115 |

| Dataset Statistics | | Dataset Statistics | |
|---|---|---|---|
| Number of Variables | 14 | Number of Variables | 15 |
| Number of Rows | 207534 | Number of Rows | 149043 |
| Missing Cells | 0 | Missing Cells | 0 |
| Missing Cells (%) | 0.0% | Missing Cells (%) | 0.0% |
| Duplicate Rows | 58491 | Duplicate Rows | 0 |
| Duplicate Rows (%) | 28.2% | Duplicate Rows (%) | 0.0% |
| Total Size in Memory | 106.9 MB | Total Size in Memory | 79.1 MB |
| Average Row Size in Memory | 540.2 B | Average Row Size in Memory | 556.5 B |
| Variable Types | Numerical: 4 Categorical: 9 GeoGraphy: 1 | Variable Types | Numerical: 5 Categorical: 9 GeoGraphy: 1 |

Figure 2. The characteristic of the UGRansome dataset

### 2.2. Stacked Autoencoder

SAEs are a type of neural network (NN) architecture that is used for feature extraction and dimension reduction in various tasks, including biometrics recognition, image recognition, natural language processing, and automatic speech recognition (Tokmak & Küçüksille, 2022). It is called "stacked" because it consists of multiple layers of autoencoders, where each layer is trained to reconstruct the output of the previous layer. The training of SAEs involves two steps: unsupervised pre-training and supervised fine-tuning. In the unsupervised pre-training step, each layer of the network is trained individually using auto-encoders, which learn internal data representations. These representations are then used to initialize the network weights and improve generalization.

In the supervised fine-tuning step, the pre-trained layers are stacked together and trained in a supervised manner using labeled data. This approach has been shown to achieve high accuracy rates in biometrics recognition tasks (Boussaad & Boucetta, 2021). In the context of automatic speech recognition, SAEs have been used to design networks for recognizing speech sounds articulated by children, achieving high accuracy rates (Wei & Zhao, 2019). SAEs can also be enhanced by incorporating data weighting techniques, which improve the robustness and discriminative power of the network (Sun et al., 2021). Additionally, SAEs have been used for automatic voice quality evaluation in call centers, achieving better correlation coefficients compared to traditional methods (L. Wang et al., 2021). In the field of intrusion detection, stacked sparse auto-encoders have been proposed for dimensionality reduction and classifiers, achieving better results compared to existing methods (Manjunatha & Gogoi, 2022). It has also been used effectively in malware detection (Rathore et al., 2018; Samaneh et al., 2022; Zhu et al., 2021). Figure 3 illustrates a SAE architecture.

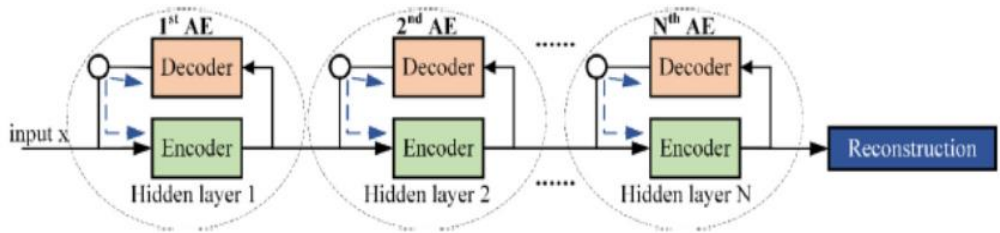

Figure 3. Structure of a SAE model (Luo et al., 2022)

### 2.3. Long Short-Term Memory

Recurrent Neural Network (RNN) is a variation of the feedforward neural network (NN). The architecture of the feedforward NN encompasses several layers, each comprised of neurons, with connections between layers proceeding unidirectionally, resulting in a sequential arrangement of layers. RNN introduces a recurrent structure within the NN by establishing connections from each node (neuron) to itself. This self-connection mechanism allows the RNN to retain previous inputs, potentially impacting the network's output (Fan et al., 2017; Naik & Mohan, 2019; Yang et al., 2021). In RNN, the inference process is like the feedforward NN completed by forward propagation. Training in RNN is done through the mechanism of backpropagation through time, updating the weights using the gradient. In RNN, the gradient for each output depends not only on the current layer but also on the previous layer. If backpropagation is continuously updated at intervals, the gradient can approach zero, leading to the vanishing gradient problem, followed by the problem of weakening gradients. Similarly, when gradients become too large, the result can grow significantly, leading to the exploding gradient problem (Metin & Karasulu, 2019). The LSTM DL algorithm was developed by Hochreiter and Schmidhuber in 1997 as a variant of the RNN model, aiming to mitigate the drawbacks associated with traditional RNN architecture (Hochreiter & Schmidhuber, 1997). Distinct from classical RNN, LSTM introduces the concept of memory cells for its nodes, enabling the linkage of prior data information to the present nodes. Each LSTM node encompasses three gating mechanisms: an input gate, a forget gate, and an output gate (Figure 4).

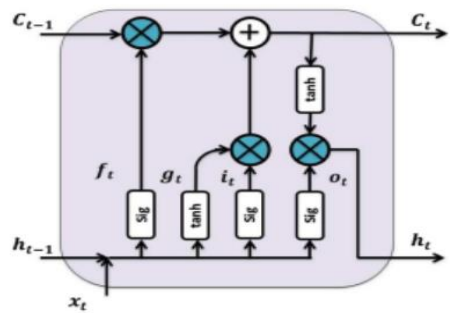

Figure 4. LSTM architecture (Smagulova & James, 2019)

### 2.4. Performance Evaluation

The evaluation of the training and testing performance of the established models was conducted by calculating accuracy, precision, recall, and F1scores. The accuracy metric, denoting the proportion of accurately classified instances used to assess the training and testing effectiveness of the formulated model, is mathematically expressed in Equation 1. The precision metric, quantifying the accuracy of positive predictions among the actual positives, is formally defined in Equation 2. The recall metric, indicating the proportion of true positive values correctly identified, is represented as Sensitivity (Recall) in Equation 3. The F1 Score, a composite metric of recall and precision, is mathematically defined as score = 2 * (precision * recall) / (precision + recall).

$$Accuracy = \frac{TP+TN}{TP+FN+TN+FP} \quad (1)$$

$$Precision = \frac{TP}{TP+FP} \quad (2)$$

$$Sensitivity = \frac{TP}{TP+FN} \quad (3)$$

In this study, the training and testing of the proposed data preprocessing, feature extraction, and classification models were conducted with the Python programming language version 3.10.12. The methodology employed for this work is explained in this section. The framework of the study is illustrated in Figure 5.

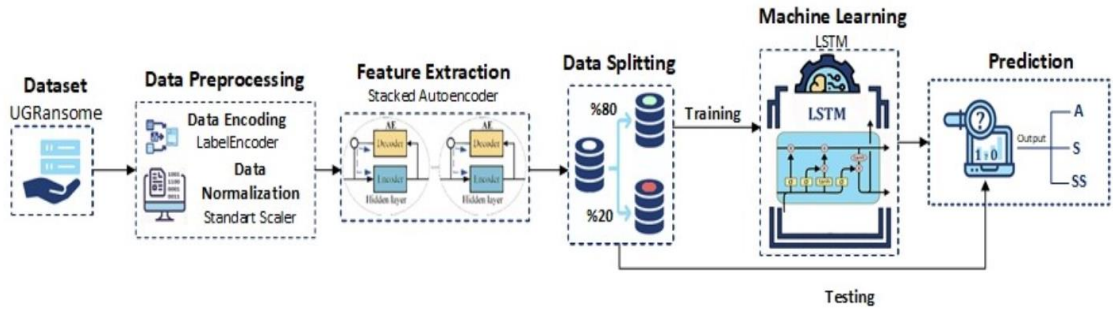

Figure 5. Proposed study model

The training and testing processes of the suggested data encoding, normalization, SAE, and LSTM model were conducted using the Google Colaboratory cloud system. This system provides ready access to numerous Python libraries and offers its services for free (*Colab*, 2023). Within the Colab platform, Nvidia CUDA technology was employed to leverage GPU acceleration for faster algorithm execution. Tasks such as file uploading, data preprocessing, setting up the data frame, and more were carried out using Python libraries including numpy, pandas, statistics, sklearn, matplotlib.pyplot, and seaborn. As for the suggested SAE and LSTM architecture, the Python TensorFlow Keras library was utilized. The suggested SAE layers and parameters (params) are given in Table 3. In the established SAE architecture, three encoders with 75, 50, and 13 layers, and three decoders with 50, 75, and 13 layers were employed. The activation function was set to relu, the optimizer parameter to Adam, the loss parameter to mse, and the epoch parameter to 50. The suggested LSTM layers and params are given in Table 4. The constructed LSTM network consists of 3 layers, each containing 168 neurons. The loss parameter was set to sparse_categorical_crossentropy, the optimizer parameter to Adam, and the epoch parameter to 400.

Table 3. SAE layers and params

| Layer (type) | Output Shape | Param # |
|---|---|---|
| input_1 (InputLayer) | (None, 13) | 0 |
| dense (Dense) | (None, 75) | 1.050 |
| dense_1 (Dense) | (None, 50) | 3.800 |
| dense_2 (Dense) | (None, 13) | 663 |
| dense_3 (Dense) | (None, 50) | 700 |
| dense_4 (Dense) | (None, 75) | 3.825 |
| dense_5 (Dense) | (None, 13) | 988 |
| Total params: | | 11.026 |

Table 4. LSTM layers and params

| Layer (type) | Output Shape | Param # |
|---|---|---|
| lstm_3 (LSTM) | (None, 168) | 122.304 |
| dense_21 (Dense) | (None, 3) | 507 |
| Total params: | | 122.811 |

## 3. RESULTS

After completing the data preprocessing and feature extraction steps, the trained LSTM model was tested, resulting in the following performance metrics: accuracy, precision, recall, and F1score were measured as 0.9849, 0.985, 0.9849, and 0.9849, respectively. These metrics are summarized in Table 5 and Figure 6.

Table 5. Performance metrics

|   | Precision | Recall | F1 score | Support |
|---|---|---|---|---|
| A | 0.971879 | 0.986131 | 0.978953 | 11320 |
| S | 0.991466 | 0.978024 | 0.984699 | 18293 |
| SS | 0.987558 | 0.994367 | 0.990951 | 11894 |
|   |   |   |   |   |
| Accuracy |   |   | 0.984918 | 41507 |
| Average | 0.985004 | 0.984918 | 0.984924 | 41507 |

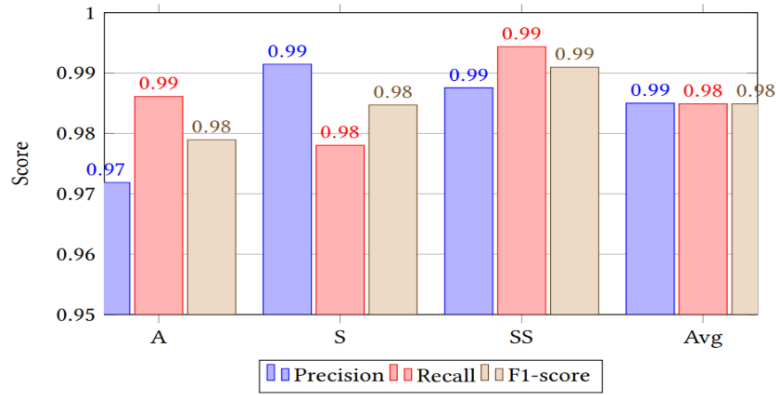

Figure 6. Visualization of the performance metrics

In summary, Table 5 and Figure 6 indicate that the SAE-based LSTM model performs well across all three attack categories—Anomaly (A), Signature (S), and Synthetic Signature (SS). It showcases high precision, recall, and F1-score values, emphasizing the model's strong predictive capabilities in identifying various types of attacks. Additionally, the balanced average scores suggest that the model generalizes effectively and consistently across different attack categories. The confusion matrix is a tool used in machine learning and classification tasks to visualize the performance of a model by presenting the number of true positive (TP), true negative (TN), false positive (FP), and false negative (FN) predictions. It is particularly useful when evaluating the accuracy of a classification algorithm. The confusion matrix showing the test results of the proposed work is shown in Figure 7.

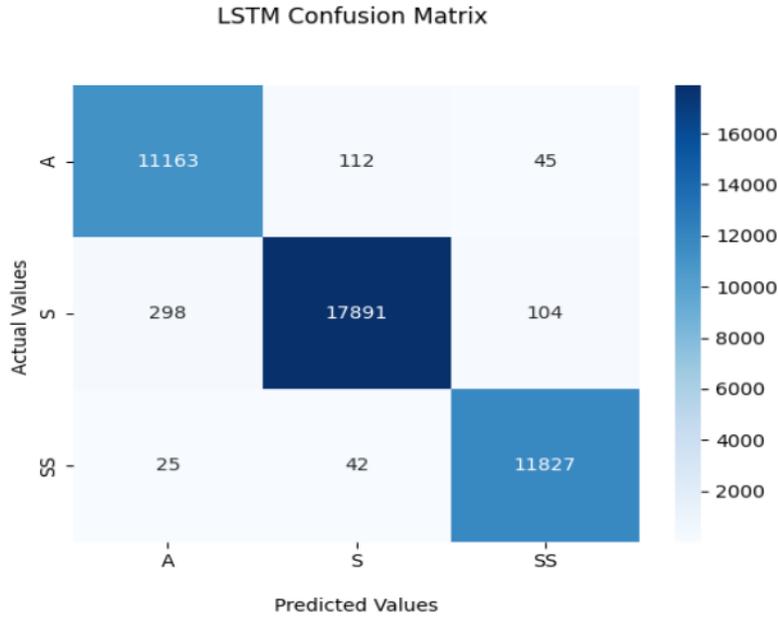

Figure 7. Confusion matrix

The SAE-based LSTM model's superior precision, recall, and F1-score in detecting signature attacks (S) compared to synthetic signature (SS) and anomaly (A) attacks signifies its effectiveness in identifying known threat patterns (Figure 7). Signature attacks are recognizable due to established patterns, and the model's adeptness in pinpointing instances aligned with these patterns is pivotal for real-time threat detection. Synthetic signature attacks involve modified or novel attack signatures, and the model's slightly lower performance could suggest difficulty in identifying unconventional or altered signatures, underscoring the challenge of evolving threat detection (Figure 7). Anomaly attacks, representing zero-day or novel threats, pose a greater detection challenge, as their lack of discernible patterns complicates identification (Figure 7). Future work in the IDS landscape can potentially use the UGRansome dataset and improve the proposed model parameters to enhance anomaly detection, adapt the model for modified signatures, implement continuous learning for emerging threats, explore ensemble approaches, and develop interpretable techniques. In essence, the proposed model's strong detection of signature attacks and potential improvements for synthetic signatures and anomalies highlight avenues for advancing IDSs.

## 4. DISCUSSION

Table 6 presents a comparative assessment of different intrusion detection studies, each employing distinct datasets and models. Our proposed methodology, centered around the UGRansome dataset, leverages the SAE-LSTM technique to achieve an impressive 98% accuracy, surpassing the performance of previous investigations. Additionally, this approach offers distinct advantages, particularly within critical infrastructure contexts. Notably, several discussed studies within the IDS literature share the common limitation of feature selection. Contrasting this, Kim et al. (2020) realized a 97% accuracy utilizing a signal autoencoder model on signal data, promising for signal processing, albeit lacking a tool diagnosis model. Zhang (2020) harnessed LSTM on malware data, achieving 81% accuracy while addressing invasive software, though grappling with dependency issues. Sun et al. (2021) embraced SAE across diverse datasets, achieving a commendable 90% accuracy with data weighting advantages, although computational time remains a concern. Blaise et al. (2020) adopted the Split Merge technique on MAWI and UCSD datasets, yielding 85% accuracy with exceptional attack detection, albeit with false positives.

Lastly, Tokmak (2022) applied deep learning to the UGRansome dataset, securing 97% accuracy with a pronounced focus on zero-day threat detection, also involving feature selection. Pertaining to zero-day attack detection, our approach not only underscores its significance but also highlights its robust accuracy and emphasis, making strides in addressing evolving threats. Considering the importance of zero-day attack detection, our approach not only underscores its significance but also demonstrates robust accuracy and a proactive stance in tackling evolving threats. To further advance cyberintelligence, future efforts could concentrate on refining feature selection techniques, addressing computational efficiency challenges, and exploring ways to integrate various methodologies to enhance overall intrusion detection capabilities.

Table 6. A comparative analysis with existing IDS studies

| Author | Year | Dataset | Model | Accuracy | Advantage | Limitation |
|---|---|---|---|---|---|---|
| Nkongolo, M., & Tokmak, M. | 2023 | UGRansome | Fuzzification | 99% | Critical Infrastructure | Feature Selection |
| Kim J et al. | 2020 | Signal data | SAE | 97% | Signal Processing | Tool diagnosis model |
| Zhang J. | 2020 | Malware | LSTM | 81% | Invasive Software | Dependency |
| Sun T et al. | 2021 | MNIST, CIFAR-10, and UCI | SAE | 90% | Data Weighting | Computational time |
| Blaise et al. | 2020 | MAWI and UCSD | Split Merge | 85% | Attack Detection | False Positive |
| Tokmak M. | 2022 | UGRansome | Deep Learning | 97% | Zero-Day Threat Detection | Feature Selection |

## 5. CONCLUSIONS

The detection and mitigation of zero-day threats have emerged as critical imperatives in the landscape of cybersecurity. Zero-day threats, by their very nature, exploit vulnerabilities that are yet unknown to software vendors and security teams, posing substantial risks to organizations and individuals. As attackers constantly evolve their techniques, the need for robust and adaptive zero-day threat detection mechanisms becomes increasingly pressing. This research endeavors to harness the potential of deep learning techniques to effectively counter the ever-evolving landscape of zero-day threats. By capitalizing on deep learning's ability to process unstructured data to provide classification and prediction analysis, we present a robust framework for zero-day threat recognition. This framework integrates the LSTM approach with SAE feature extraction. The encouraging results obtained from our detection system underscore its significant effectiveness, thus offering valuable guidance and inspiration for forthcoming research pursuits in this field. In conclusion, the fight against zero-day threats demands a multi-faceted approach that integrates cutting-edge technology, collaborative efforts, and robust risk management practices. While machine learning and deep learning are potent tools, they must be complemented by human expertise to effectively counteract the sophistication of modern cyber threats. As the cybersecurity landscape continues to evolve, the ability to detect and mitigate zero-day threats will be a defining factor in ensuring the security and stability of digital ecosystems.

**REFERENCES**

Blaise, A., Bouet, M., Conan, V., & Secci, S. (2020). Detection of zero-day attacks: An unsupervised port-based approach. *Computer Networks*, *180*, 107391.

Boussaad, L., & Boucetta, A. (2021). Stacked Auto-Encoders Based Biometrics Recognition. *2021 International Conference on Recent Advances in Mathematics and Informatics (ICRAMI)*, 1–6.

Chatterjee, S., Dey, D., & Munshi, S. (2019). Morphological, texture and auto-encoder based feature extraction techniques for skin disease classification. *2019 IEEE 16th India Council International Conference (INDICON)*, 1–4.

Fang, Y., Huang, C., Liu, L., & Xue, M. (2018). Research on Malicious JavaScript Detection Technology Based on LSTM. *IEEE Access*, *6*, 59118–59125. https://doi.org/10.1109/ACCESS.2018.2874098

*Google Colaboratory*. (2023). Colaboratory. https://colab.research. google.com/

Kim, J., Lee, H., Jeon, J. W., Kim, J. M., Lee, H. U., & Kim, S. (2020). Stacked auto-encoder based CNC tool diagnosis using discrete wavelet transform feature extraction. *Processes*, *8*(4), 456.

Kong, X., Lin, R., & Zou, H. (2020). Feature extraction of load curve based on autoencoder network. *2020 IEEE 20th International Conference on Communication Technology (ICCT)*, 1452–1456.

Kumar, V., & Sinha, D. (2021). A robust intelligent zero-day cyber-attack detection technique. *Complex & Intelligent Systems*, *7*(5), 2211–2234. https://doi.org/10.1007/s40747-021-00396-9

Luo, S., Huang, X., Wang, Y., Luo, R., & Zhou, Q. (2022). Transfer learning based on improved stacked autoencoder for bearing fault diagnosis. *Knowledge-Based Systems*, *256*, 109846.

Manjunatha, B. A., & Gogoi, P. (2022). An Improved Stacked Sparse Auto-Encoder Method for Network Intrusion Detection. *Emerging Research in Computing, Information, Communication and Applications: ERCICA 2020, Volume 1*, 103–118.

Millar, S., McLaughlin, N., del Rincon, J. M., & Miller, P. (2021). Multi-view deep learning for zero-day Android malware detection. *Journal of Information Security and Applications*, *58*, 102718.

Nkongolo, M., & Tokmak, M. (2023). Zero-Day Threats Detection for Critical Infrastructures. In A. Gerber & M. Coetzee (Eds.), *South African Institute of Computer Scientists and Information Technologists* (pp. 32–47). Springer Nature Switzerland. https://doi.org/10.1007/978-3-031-39652-6_3

Nkongolo, M., Van Deventer, J. P., & Kasongo, S. M. (2021). Ugransome1819: A novel dataset for anomaly detection and zero-day threats. *Information*, *12*(10), 405.

Nkongolo, M., van Deventer, J. P., & Kasongo, S. M. (2022). The Application of Cyclostationary Malware Detection Using Boruta and PCA. In *Computer Networks and Inventive Communication Technologies* (pp. 547–562). Springer.

Nkongolo, M., van Deventer, J. P., Kasongo, S. M., & van der Walt, W. (2022). Classifying social media using deep packet inspection data. In *Inventive Communication and Computational Technologies: Proceedings of ICICCT 2022* (pp. 543-557). Singapore: Springer Nature Singapore.

Rathore, H., Agarwal, S., Sahay, S. K., & Sewak, M. (2018). Malware detection using machine learning and deep learning. *Big Data Analytics: 6th International Conference, BDA 2018, Warangal, India, December 18–21, 2018, Proceedings 6*, 402–411.

Roberts, C., & Nair, M. (2018). Arbitrary discrete sequence anomaly detection with zero boundary LSTM. *ArXiv Preprint ArXiv:1803.02395*.

Samaneh, M., Dima, A., & Ghorbani, A. A. (2022). Effective and Efficient Hybrid Android Malware Classification Using Pseudo-Label Stacked Auto-Encoder. *Journal of Network and Systems Management*, *30*(1).

Sarhan, M., Layeghy, S., Gallagher, M., & Portmann, M. (2023). From zero-shot machine learning to zero-day attack detection. *International Journal of Information Security*, *22*(4), 947–959. https://doi.org/10.1007/s10207-023-00676-0

Sewak, M., Sahay, S. K., & Rathore, H. (2021). LSTM Hyper-Parameter Selection for Malware Detection: Interaction Effects and Hierarchical Selection Approach. *2021 International Joint Conference on Neural Networks (IJCNN)*, 1–9. https://doi.org/10.1109/IJCNN52387.2021.9533323

Smagulova, K., & James, A. P. (2019). A survey on LSTM memristive neural network architectures and applications. *The European Physical Journal Special Topics*, *228*(10), 2313–2324.

Sun, T., Ding, S., & Xu, X. (2021). An iterative stacked weighted auto-encoder. *Soft Computing*, *25*(6), 4833–4843. https://doi.org/10.1007/s00500-020-05490-7

Thomas, V. E., Ugwu, C., & Onyejegbu, L. (2021). Comparative Analysis of Dimensionality Reduction Techniques on Datasets for Zero-Day Attack Vulnerability. *Journal of Software Engineering and Simulation*, *7*(8), 48–56.

Tokmak, M. (2022). Deep forest approach for zero-day attacks detection. In S. Tasdemir & A. O. Ozkan (Eds.), *Innovations and Technologies in Engineering* (pp. 44–55). Eğitim Yayınevi.

Tokmak, M., & Küçüksille, E. U. (2022). Comparative Analysis of Dimension Reduction and Classification Using Cardiotocography Data. In M. Karaboyacı & A. Demirçalı (Eds.), *Versatile Multidisciplinary Engineering Research* (pp. 149–164). SRA Academic Publishing.

Wang, L., Wang, Z., Zhao, G., Su, Y., Zhao, J., & Wang, L. (2021). Automatic voice quality evaluation method of IVR service in call center based on Stacked Auto Encoder. *IOP Conference Series: Earth and Environmental Science*, *827*(1), 012021.

Wang, T., Ng, W. W., Li, W., & Kwong, S. (2021). Broad Autoencoder Features Learning for Classification Problem. *International Journal of Cognitive Informatics and Natural Intelligence (IJCINI)*, *15*(4), 1–15.

Wang, Y., Yang, H., Yuan, X., Shardt, Y. A., Yang, C., & Gui, W. (2020). Deep learning for fault-relevant feature extraction and fault classification with stacked supervised auto-encoder. *Journal of Process Control*, *92*, 79–89.

Wei, P., & Zhao, Y. (2019). A novel speech emotion recognition algorithm based on wavelet kernel sparse classifier in stacked deep auto-encoder model. *Personal and Ubiquitous Computing*, *23*(3), 521–529. https://doi.org/10.1007/s00779-019-01246-9

Xie, Q., Wang, Y., & Qin, Z. (2020). Malware Family Classification using LSTM with Attention. *2020 13th International Congress on Image and Signal Processing, BioMedical Engineering and Informatics (CISP-BMEI)*, 966–970. https://doi.org/10.1109/CISP-BMEI51763.2020.9263499

Xie, W., Xu, S., Zou, S., & Xi, J. (2020). A System-call Behavior Language System for Malware Detection Using A Sensitivity-based LSTM Model. *Proceedings of the 2020 3rd International Conference on Computer Science and Software Engineering*, 112–118. https://doi.org/10.1145/3403746.3403914

Zhang, J. (2020). DeepMal: A CNN-LSTM Model for Malware Detection Based on Dynamic Semantic Behaviours. *2020 International Conference on Computer Information and Big Data Applications (CIBDA)*, 313–316. https://doi.org/10.1109/CIBDA50819.2020.00077

Zhu, H.-J., Wang, L.-M., Zhong, S., Li, Y., & Sheng, V. S. (2021). A hybrid deep network framework for android malware detection. *IEEE Transactions on Knowledge and Data Engineering*, *34*(12), 5558–5570.